\documentclass[12pt]{article}
\usepackage{graphicx}
\usepackage{latexsym,amsfonts, amssymb,epsfig}
\usepackage{amssymb,amsmath,amsthm}
\usepackage[english]{babel}

\begin{document}

\begin{center}
{\Large\bf On the localized wave patterns supported by
convection--reaction--diffusion equation}\footnote{The research was
supported by the AGH local grant}

\vspace{10mm}

{\it {\large\bf V. Vladimirov\footnote{{\it E-mail address:}
vsevolod.vladimirov@gmai.com } and Cz. M\c{a}czka}
\\
 \vspace{5mm}

Faculty of Applied Mathematics \\
  University of Science and Technology\\
Mickiewicz Avenue 30, 30-059 Krak\'{o}w, Poland \\
 [2ex] }

 \end{center}

\vspace{10mm}

{ \footnotesize {\bf Abstract.  } A set of traveling wave solution
to convection-reaction-diffusion equation is studied by means of
methods of local nonlinear analysis and numerical simulation. It is
shown the existence of compactly supported solutions as well as
solitary waves within this family for wide range of parameter
values.

 }

 \vspace{3mm}


\vspace{3mm}


\vspace{3mm}

\section{ Introduction }

 We consider evolutionary equation
\begin{equation}\label{GBE1}
u_t+u\,u_x-\kappa\,\left(u^n\,u_x
\right)_x=\left(u-U_1\right)\,\varphi(u),
\end{equation}
where $\kappa,\,\,\,U_1$ are positive constants, $\varphi(u)$ is a
smooth function nullifying at the origin and maintaining a constant
sign within the set $\left(0,\,k\,U_1\right)$ for some $k>1$.
Equation (\ref{GBE1}) referred to as convection--reaction--diffusion
equation is used for simulating transport phenomena in active media.
Due to its practical applicability and a number of unusual features,
equation (\ref{GBE1}) was intensely studied in recent decades
\cite{Tang,Korman,Lou,Kersner,Nekhamkina,Smoller,Cherniha1,BarYur}.

Present investigations are mainly devoted to qualitative and
numerical study of the family of traveling wave (TW) solution to
equation (\ref{GBE1}). Our aim is to show that under certain
conditions the set of TW solutions contains periodic regimes,
solitary waves and compactons  with special emphasis put on the
existence of compactons. Solitary wave solutions (or solitons) are
exponentially localized wave packs moving with constant velocity
without change of their shape. They are mostly associated with the
famous Korteveg de Vries (KdV) equation and members of KdV hierarchy
having the form
\begin{equation}\label{KdVeq}
K(m)=u_t+u^m\,u_x+u_{xxx}=0, \qquad m \ge1.
\end{equation}
For $m=1$ solitary wave solution to KdV is as follows \cite{Dodd}:
\[
u(t,\,x)=U(\xi)\equiv
U(x-V\,t)=\frac{3\,V}{\cosh\left(\frac{\sqrt{V}}{2}\,\xi\right)},
\]
where $V$ stands for the wave pack velocity.

In 1993 Philip Rosenau and John Hyman put forward the following
generalization to KdV hierarchy \cite{Ros_93}:
\begin{equation}\label{KdVeq}
K(m,\,n)=u_t+\left(u^m\right)_x+\left(u^n\right)_{xxx}=0, \qquad
m,\,n \ge 2.
\end{equation}
Modification in the higher-derivative term causes that members of
$K(m,\,n)$ hierarchy possess solutions with compact support. For
$m=n=2$ compactly supported solution (or {\it compacton}) has the
form
\[
 u(t,\,x)=U(\xi)\equiv
U(x-V\,t)=\left\{\begin{array}{c}
\frac{4\,V}{3}\,\cos^2\left[\frac{x-V\,t}{4} \right] \quad \mbox{if}
\quad \left|\frac{x-V\,t}{4}\right| \\
0 \quad \mbox{otherwise}.  \end{array} \right.
\]
Both solitons and compactons are the TW solutions, depending in fact
on a single variable $\xi=x-V\,t$, are described by ODEs appearing
when the TW ansatz $u(t,\,x)=U(\xi)$ is inserted into the source
equations (for details see e.g. \cite{Vlad_08}). Therefore it is
possibile to give a clear geometric interpretation of solitons and
compactons. To both of them correspond homoclinic trajectories in
the phase space of the factorized equations. The main difference
between
 the solitary wave and compacton is that the first one is nonzero for any
 $\xi$ while  the last one is nonzero
 only upon some compact set. This is because the soliton corresponds to
 the homoclinic loop usually bi-asymptotic
 to a simple saddle. Hence the ''time'' needed to penetrate the close loop
 is infinite. Compacton corresponds to the homoclinic loop
 bi-asymptotic to a topological saddle lying on a singular line. A
 consequence of this is that the corresponding vector field does not
 tend to zero as the homoclinic trajectory approaches the stationary
 point, hence the ''time'' required to penetrate the trajectory remains
 finite. Let us note that the compacton in fact is a conjunction of
 the nonzero part corresponding to the homoclinic loop and the
 trivial constant solution represented by the saddle point \cite{Vlad_08}.

For the sake of brevity we maintain the notation traditionally used
in more specific sense. Thus, soliton is usually associated with the
localized invariant solution to any completely integrable PDE,
possessing a number of unusual features \cite{Dodd}. Some of these
features are inherited by compactons as well \cite{Ros_93,
Pik_Ros1,Pik_Ros2}. Here we maintain this notion to those solutions
to (\ref{GBE1}), which manifest similar geometric features as
''true'' wave patterns known under these names, not pretending that
they inherit all features of their famous precursors.

The structure of the article is following. In section 2 we present
the dynamical system describing the TW solutions to (\ref{GBE1}) and
make its qualitative analysis, obtaining  conditions that contribute
to the homoclinic loop appearance. At the end of this section we
formulate the conditions enabling to make a distinction between
compactons and solitons. Conditions formulated in section 2
contribute but not guarantee the homoclinic loop appearance. The
following section contains the results of numerical simulations
revealing that expected scenarios of solitons' and compactons'
appearance really take place. In section 4 we perform a brief
discussion of the results obtained and outline directions of further
investigations.

\section{Factorized system and its qualitative analysis}

\subsection{Statement of the problem}

We are going to analyze the set of  TW solutions to (\ref{GBE1}),
described by the following formula:
\begin{equation}\label{twans}
u(t,\,x)=U(\xi)\equiv U\left(x-V\,t\right).
\end{equation}
Inserting ansatz (\ref{twans}) into the GBE one can obtain, after
some manipulation, the following dynamical system:
\begin{eqnarray}\label{factors1}
\Delta(U)\,\dot U=\Delta(U)\,W,    \\
\Delta(U)\,\dot
W=\varphi(U)\left(U_1-U\right)-\kappa\,n\,U^{n-1}\,W^2+\left(U-V\right)\,W,
\nonumber
\end{eqnarray}
where $\Delta(U)=\kappa\,U^n$. We are going to formulate the
conditions contributing to soliton-like and compacton-like solutions
to equation (\ref{GBE1}) by analyzing the factorized system
(\ref{factors1}). Let us note that in the case of both KdV and
$K(m,\,n)$ hierarchies' members the procedure of factorization lead
to the Hamiltonian dynamical systems for which distinguishing the
presence of homoclinic loops is more or less trivial \cite{Vlad_08}.
The situation with system (\ref{factors1}) is not so simple since it
is not Hamiltonian. And we cannot expect that the homoclinic loops
will form a one-parametric family as this is the case with the
Hamiltonian systems. Contrary, the homoclinic loop in our case can
appear as a result of one or several successive bifurcations taking
place at specific values of the parameters. Let us stress that
existence of the homoclinic loop corresponding to compacton is
possible due to the presence of the factor $\Delta(U)$ at the LHS of
system (\ref{factors1}). Since the function $\Delta(U)$ contains the
origin, which is the stationary point of system (\ref{factors1}),
then the natural way of proceedings is following. We state the
condition that guarantee the appearance of the stable limit cycle in
proximity of the stationary point $(U_1,\,0)$, and ensure that
simultaneously the stationary point $(0,\,0)$ is a saddle and remain
so as the parameter of bifurcation is changed. By  proper change of
the bifurcation parameter we can cause the growth of size of the
limit cycle and in presence of near-by saddle it finally could give
way to the homoclinic bifurcation. Since the latter bifurcation is
nonlocal, we cannot trace its appearance using the methods of local
nonlinear analysis, as this is the case with Andronov-Hopf
bifurcation. Therefore on the final step we are forced to resort to
numerical simulations. And of course answer the question on whether
the homoclinic loop corresponds to compacton-like solution or not
needs special treatment based upon the asymptotic analysis. This and
other issues are analyzed in the following subsections.

\subsection{Andronov-Hopf bifurcation in system (\ref{factors1})}

  To formulate the
conditions which guarantee the stable limit cycle appearance in
vicinity of the stationary point $(U_1,\,0)$, let us consider the
Jacobi matrix
\[
J_1=
\left(\begin{array}{cc} 0 & \Delta(U_1) \\
-{\varphi}(U_1) & U_1-V.
\end{array}
\right).
\]
 corresponding to this point. In order that  $(U_1,\,0)$ be a center,
 the eigenvalues of $J_1$ should be pure imaginary and this is so
 when the following conditions are fulfilled:
\begin{eqnarray}
\mbox{Trace}\,{J_1}=U_1-V=0, \label{tr_limc} \\
\mbox{Det}\,{J_1}=\Delta(U_1) \varphi(U_1)>0 \label{det_limc}
\end{eqnarray}
The first condition immediately gives us the critical value of the
wave pack velocity $V_{cr}=U_1.$ The second one is equivalent to the
statement that  $\varphi(U_1)$ is positive.

The next thing we are going to do is a study of the stability of
limit cycle appearing. As is well known \cite{Has,GH}, this is  the
real part of the first Floquet index $\Re{C_1}$ that determines
whether the cycle is stable or not.

For $\Delta(U_1)=\kappa U_1^n>0$ condition  $\Re{C_1}<0$ assures
that the limit cycle, appearing in system when $V<V_{cr}$, will be
stable.

To obtain the expression for $\Re{C_1}$, the standard  formula
contained e.g. in \cite{GH} could be applied providing that system
in vicinity of the center is presented in the form
\begin{eqnarray}\label{cpf_1}
\left(\begin{array}{c}\dot z_1
\\\dot z_2 \end{array} \right)=\left(\begin{array}{lc} 0  & -\Omega \\ \Omega
& 0
\end{array} \right)\cdot \left(\begin{array}{c}z_1 \\z_2
\end{array}\right)+\left(\begin{array}{c}F(z_1,z_2)
\\G(z_1,z_2) \end{array}\right),
\end{eqnarray}
where $\Omega=\sqrt{\mu\cdot \nu},$ $\mu=\Delta(U_1)$,
$\nu=\varphi(U_1)$, $F(z_1,z_2)$ and $G(z_1,z_2)$ contain all
nonlinear terms. In this  (canonical) representation the real part
of the Floquet index $\Re\,C_1$ is expressed by the following
formula \cite{GH}:
\begin{eqnarray}\label{floqind}
16\,
\Re\,C_1=F_{111}+F_{122}+G_{112}+G_{222}+\frac{1}{\Omega}\left\{F_{12}\left(F_{11}+F_{22}\right)-
\right. \nonumber \\\left. -G_{12}\,\left(G_{11}+G_{22}\right)
-F_{11}\,G_{11}+F_{22}\,G_{22}  \right\}.
\end{eqnarray}
Here $F_{ijk},\,\,F_{ij}$ stand for the coefficients of the
function's $F(z_1,z_2)$ monomials $z_i\,z_j\,z_k$, $z_i\,z_j$
correspondingly. Similarly, indices $G_{ijk},\,\,G_{ij}$ correspond
to the function's $G(z_1,z_2)$ monomials' coefficients.

Assuming that relations (\ref{tr_limc})--(\ref{det_limc}) are
satisfied, let us rewrite system (\ref{factors1} ) in coordinates
$y_1=U-U_1,\,\,y_2=W$:
\begin{equation}\label{auxcommon}
\Delta(U) \frac{d}{d\,\xi}\left( \begin{array}{c} y_1 \\y_2
\end{array} \right)=\left( \begin{array}{lc} 0 & \mu \\ -\,\nu &
0
\end{array} \right)\left( \begin{array}{c} y_1 \\y_2
\end{array} \right)+\left( \begin{array}{c} \Phi_1(y_1,\,y_2) \\\Phi_2(y_1,\,y_2)
\end{array} \right),
\end{equation}
where $\Phi_1(y_1,\,y_2), \,\,\Phi_2(y_1,\,y_2)$ are nonlinear terms
which are as follows:
\begin{eqnarray}
\Phi_1(y_1,\,y_2)=\kappa n U_1^{n-2}\,y_1\,y_2
\left(U_1+\frac{n-1}{2}y_1 \right)+O\left(|y_i|^4  \right),
\qquad\qquad\qquad
\\
\Phi_2(y_1,\,y_2)=-\left\{\kappa\,U_1^{n-1}y_2^2+n(n-1)\kappa\,U_1^{n-2}y_1\,y_2^2+\qquad\qquad\qquad\qquad\qquad
\right.
\\ \left. +y_1\left[\dot{\varphi}\left(
U_1\right)\,y_1+\frac{1}{2}\ddot{\varphi}\left(
U_1\right)\,y_1^2-y_2 \right] \right\}+O\left(|y_i|^4\right).
\nonumber
\end{eqnarray}

A passage to canonical variables $(z_1,\,z_2)$ can be attained by
means of  transformation
\begin{eqnarray*}
\left(\begin{array}{c}z_1 \\z_2 \end{array}
\right)=\left(\begin{array}{lc} -\sqrt{\nu} & 0 \\ 0 & \sqrt{\mu}
\end{array} \right)\cdot \left(\begin{array}{c}y_1 \\y_2 \end{array}\right),
\end{eqnarray*}
This gives us the canonical system (\ref{cpf_1}) with
\begin{eqnarray} F(z_1,z_2)=
\frac{\kappa\,n\,U_1^{n-2}}
{\Omega }z_1\,z_2\,\left[\sqrt{\nu}\, U_1 -\frac{n-1}{2}z_1 \right]+O(|z_i|^4),\label{FG_canon}\\
G(z_1,z_2)=-\frac{\mu}{2\,\nu\Omega}z_1^2\left[2\,\dot
\varphi(U_1)\sqrt{\nu}-\,{\ddot \varphi(U_1)}z_1
\right]-\,\frac{\sqrt{\mu}}{\Omega}{z_1 \,z_2}+ \qquad\qquad\qquad
\nonumber
\\ +\frac{\kappa\, n\, U_1^{n-2}}{\Omega} z_2^2 \left[
(n-1)\, z_1-U_1\sqrt{\nu} \right]+O(|z_i|^4).\qquad\qquad\qquad
\nonumber
\end{eqnarray}

 Analysis of the formulae (\ref{floqind}), (\ref{FG_canon}) shows that the function $F(z_1,z_2)$ does not
contribute to the real part of the Floquet index, which in our case
is  expressed as follows:
\[
\Re{ C_1}=-\frac{1}{16\,\Omega}\,G_{12}\,\left(G_{11}+G_{22}
\right)=-\,\frac{1}{16\,\Omega^2\,\varphi(U_1)}
\left\{\Delta(U_1)\,\dot{\varphi}(U_1)+ \kappa\, n
\varphi(U_1)\,U_1^{n-1} \right\}.
\]

So the following statement holds.

{\bf Theorem 1.} {\it
   If the inequality
\begin{equation}\label{ineq}
U_1\,\dot{\varphi}(U_1)+  n \varphi(U_1)\,>\,0
\end{equation}
is fulfilled then in some vicinity of the critical value of the wave
pack velocity $V_{cr}=U_1$ a stable limit cycle exists. }

 \subsection{Study of the stationary point
 $(0,\,0)$}\label{asymptorigin}

 From now on we shall investigate system with
 $\varphi(U)=U^m$:
\begin{eqnarray}\label{factors2}
\Delta(U)\,\dot U=\Delta(U)\,W,    \\
\Delta(U)\,\dot
W=\varphi(U)\left(U_1-U\right)-\kappa\,n\,U^{n-1}\,W^2+\left(U-V\right)\,W,
\nonumber
\end{eqnarray}

 Our next step is to state the conditions assuring that the
stationary point $(0,0)$ is a topological saddle or at least
contains a saddle sector in the right half-plane. The standard
theory \cite{AndrLeont} can be applied for this purpose. Our system
can be written down in the form
\begin{equation}\label{precan}
\frac{d}{d\,T}\left(\begin{array}{c} U \\W
\end{array}\right)=\left(\begin{array}{lc} 0 & 0 \\U_1\,\dot\varphi(0) & -
V \end{array}\right)\left(\begin{array}{c} U \\W
\end{array}\right)+\mbox{nonl.\,\,terms},
\end{equation}
where $\frac{d}{d\,T}=\kappa\,U^n\,\frac{d}{d\,\xi}$ and since the
trace of its linearization matrix is nonzero then the
 analysis prescribed in Chapter IX of \cite{AndrLeont} for this type of
 a complex stationary points is the following.
\begin{enumerate}

\item
Find the change of variables
$\left(U,\,W,\,T\right)\mapsto\left(x,\,y,\,\tau\right) $ enabling
to write down system (\ref{precan}) in the standard form
\[
\frac{d\,x}{d\,\tau}=P_2\left(x,\,y\right),
\]
\[
\frac{d\,y}{d\,\tau}=y+Q_2\left(x,\,y \right),
\]
where $P_2\left(x,\,y\right)$, $Q_2\left(x,\,y \right)$ are
polynomials of degree 2 or higher.

\item

Solve the equation $y+Q_2\left(x,\,y \right)=0$ with respect to y,
presenting result in the form of the asymptotic decomposition
$y=\varphi(x)=\alpha_1\,x^{\mu_1}+\alpha_2\,x^{\mu_2}+...$.

\item

Find the asymptotic decomposition
\[
P_2\left(x,\,\varphi(x)\right)=\Delta_m\,x^{m}+....
\]

\item
Depending on the values of $m$ and the sign of $\Delta_m$ select the
type of the complex stationary point using the theorem 65 from
\cite{AndrLeont}.

\item

Return to the original variables $\left(U,\,W,\,T\right)$ and
analyze on whether the geometry of the problem allows for the
homoclinic bifurcation appearance.

\end{enumerate}

So let us present results obtained for system (\ref{factors2}). The
case $m>1$  is most simple for analyzing since the canonical system
is obtained by the formal change $\left(U,\,W\right)\mapsto
\left(x,\,y\right)  $ and passage to new independent variable
$\tau=-V\,T$. As a result we get the following system:
\begin{eqnarray}
\frac{d\,x}{d\,\tau}=-\frac{\kappa}{V}x^n\,y,=P_2(x,\,y), \label{ALaux1}\\
\frac{d\,y}{d\,\tau}=y-\frac{1}{V}\left\{x y
-\kappa\,n\,x^{n-1}\,y^2-x^m\,\left(  x-U_1\right)
\right\}=y+Q_2(x,\,y).  \nonumber
\end{eqnarray}
Presenting $y$ in the form of series $y=\alpha_1\,x^\mu_1+...$ and
solving the equation $y+Q_2(x,\,y)=0,$ we obtain
$y=\varphi(x)=\frac{U_1}{V}x^m+...$. Inserting function $\varphi(x)$
into the RHS of the first equation we get
\[
P_2(x,\,\varphi(x))=\Delta_{n+m}
x^{n+m}+...=-\kappa\,U_1\,x^{n+m}+....
\]

Case $m=1$ is different from the previous one. Passage to canonical
system is attained by means of the following transformation:
\[
x=U, \quad y=W-\frac{U_1}{V}\,U, \quad \tau=-V\,T.
\]
In new variables system read as follows:
\begin{eqnarray}
\frac{d\,x}{d\,\tau}=-\frac{\kappa}{V}x^n\,\left(\frac{U_1}{V}x+y  \right)=P_2(x,\,y),
\qquad\label{ALaux2}\\
\frac{d\,y}{d\,\tau}=y-\frac{1}{V}\left\{x \left(\frac{U_1}{V}x+y
\right) -\kappa\,n\,x^{n-1}\,\left(\frac{U_1}{V}x+y
\right)^2-x^2+U_1\,P_2(x,\,y)\right\}= \nonumber \\ =y+Q_2(x,\,y).
\nonumber
\end{eqnarray}
Solving  equation $y+Q_2(x,\,y)=0,$ we get
$y=\varphi(x)=a_2\,x^2+...$, where index  $a_2$ depends upon $n$
Inserting function $\varphi(x)$ into the RHS of the first equation
we get
\[
P_2(x,\,\varphi(x))=\Delta_{n+1}
x^{n+1}+...=-\kappa\,U_1\,x^{n+1}+....
\]

\begin{figure}
\includegraphics[width=3.75 in, height=2.25 in]{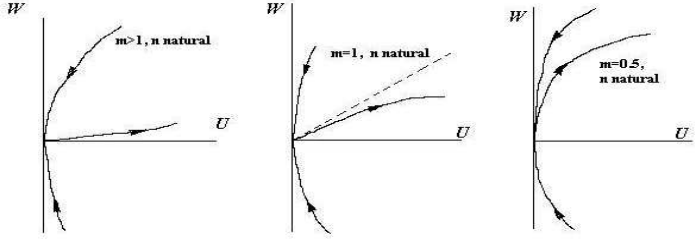}
\caption{Phase portraits in vicinity of $(0,\,0) $ for different
values of the parameter $m$ }\label{fig:topsaddle}
\end{figure}

The last case we are going to analyze is that with $m=\frac{1}{2}$.
The motivation of such a choice will be clear later on. In order
that the analytical theory be applicable we make a change variables
$Z=\sqrt{U}$. After that we get the system
\begin{eqnarray}
\frac{d\,Z}{d\,T}=-\frac{\kappa}{2}Z^{2\,n-1}\,W,\label{ALaux} \\
\frac{d\,W}{d\,T}=W\left(Z^2-V\right)-\kappa\,n\,Z^{2(n-1)}W^2+Z\,\left(U_1-Z^2\right).
\nonumber
\end{eqnarray}
Passage to the canonical variables is performed by the change of
variables
\[
x=Z, \quad y=W-\frac{U_1}{V}Z, \quad \tau=-D\,T.
\]
The system resulting from this  will be as follows:
\begin{eqnarray}
\frac{d\,x}{d\,\tau}=-\frac{\kappa}{2\,V}x^{2n-1}\,\left(\frac{U_1}{V}x+y
\right)=P_2(x,\,y), \label{ALaux3} \\
\frac{d\,y}{d\,\tau}=y-\frac{1}{V}\left\{x^2 \left(\frac{U_1}{V}x+y
\right) -\kappa\,n\,x^{2(n-1)}\,\left(\frac{U_1}{V}x+y
\right)^2-x^3+U_1\,P_2(x,\,y)\right\}= \nonumber \\=y+Q_2(x,\,y).
\nonumber
\end{eqnarray}
Solving first equation $y+Q_2(x,\,y)$ with respect to $y$ and
inserting the function $\varphi(x)$ obtained this way into $P_2$ we
get the decomposition
\[
P_2(x,\,\varphi(x))=\Delta_{2n}\,x^{2n}+...=-\frac{\kappa}{2\,V}x^{2n}+...
\]

Using  the classification given in \cite{AndrLeont} we are now able
to formulate the following result.

\newpage

{\bf Theorem 2.} {\it
 \begin{enumerate} \item  If  $m+n$ is an odd
natural number then the origin of system (\ref{ALaux1}) is a
topological saddle having a pair of separatrices tangent to the
vertical axis and the other pair tangent to the horizontal axis when
$m> 1$ and to the line $W=\frac{U_1}{V}\,U$ when $m=1$. For even
$m+n$ the stationary point is a saddle-node with two saddle sectors
lying in the right half-plane. Two of its three separatrices are
tangent to the vertical axis while the remaining one is tangent to
the horizontal axis in case when $m> 1$ and and to the line
$W=\frac{U_1}{V}\,U$ when $m=1$. The nodal sector lying in the left
half-plane is unstable.
\item  If $n$ is an odd number then the origin of system
corresponding to (\ref{ALaux2}) is a topological saddle identical
with that of the previous case. For even $n$ the stationary point is
a saddle-node identical with that of the previous case
\item For any $n \ge 1$ the origin of system (\ref{ALaux3}) is a
saddle-node geometrically identical with that from the previous two
cases.
\end{enumerate}
}


The crucial fact appearing from this analysis is that the stationary
points $(0,\,0)$ of the canonical systems
(\ref{ALaux1})--(\ref{ALaux3}) depending on the values of the
parameters $m,\,n$ are either  saddles or saddle-nodes  with the
saddle sectors placed at the right half-space. The return to the
original coordinates does not cause the change the position of the
saddle sectors but changes the orientation of vector fields and the
angle at which the outgoing separatrice leaves the stationary point.
The local phase portraits corresponding to three distinct cases are
shown on figure \ref{fig:topsaddle} reconstructed on the basis if
the analysis of relation between (\ref{ALaux1})--(\ref{ALaux3}) and
system (\ref{factors2}).

Before we start to discuss the results of numerical study of system
(\ref{factors2}), let us consider the question on when the
presumably appearing homoclinic loop corresponds to the
compacton-like solution to the source equation (\ref{GBE1}). To
answer this question we are going to find the solution
$W(U)=\alpha_1\,U^{\mu_1}+\alpha_2\,U^{\mu_2}+...$  of equation
\begin{equation}\label{asymptd}
\kappa\,U^n\,W\,\frac{d\,W}{d\,U}=G\equiv
U^m\,(U_1-U)-\kappa\,n\,U^{n-1}\,W^2-\left(V-U\right)\,W,
\end{equation}
(equivalent to system (\ref{factors2})) up to the given order and
next estimate the limit of the vertical component of vector field
\begin{equation}\label{limVF}
K=\lim_{U \to 0}\frac{G\left(U, W(U)\right)}{\kappa\,U^n}.
\end{equation}
Homoclinic loop will correspond to comactly-supported solution if K
is finite or infinite and also if it tends to zero not quicker than
$U^\mu,\,\,\mbox{where}\,\,1/2<\mu<1.$

Equation (\ref{asymptd}) can be re-written as follows:
\begin{eqnarray}\label{asymptd}
\kappa\,U^{n+2\,\mu_1-1} \left(
\alpha_1+\alpha_2\,U^{\mu_2-\mu_1}+...\right)\,\left(
\mu_1\,\alpha_1+\mu_2\,\alpha_2\,U^{\mu_2-\mu_1}+...\right)= \nonumber \\
= U^{1+\mu_1}\left(\alpha_1+\alpha_2\,U^{\mu_2-\mu_1}+... \right)-
V\,\left(\alpha_1+\alpha_2\,U^{\mu_2-\mu_1}+... \right)-
\\
 -\kappa\,n\,U^{n+2\,\mu_1-1}\left(\alpha_1^2+2\alpha_1\alpha_2\,U^{\mu_2-\mu_2}+....\right)-
 U^{m+1}+U_1\,U^m.  \nonumber
\end{eqnarray}
The procedure of solving (\ref{asymptd}) is pure algebraic: we
collect the coefficient of different powers of $U$ end equalize them
to zero. The lowest power in the RHS is either $U^{\mu_1}$ or
$U^{\mu_1}$ or $U^{n+2\,\mu_1-1}$. The number ${n+2\,\mu_1-1}$
cannot be less or equal to $\mu_1$ because it involves the
inequality $0<\mu_1 \le 1-n$ which is impossible for any natural
$n$. On the other hand if ${n+2\,\mu_1-1} \le m$ then $\mu_1$
becomes an ''orphan'' and $\alpha_1$ should be nullified.

So $U^{n+2\,\mu_1-1}$ cannot be the lowest monomial. From this
immediately appears that the only choice leading to the nontrivial
solution is $\mu_1=m$. Next important observation is such that the
first nonzero monomial at the RHS should have as a counterpart at
the LHS the least monomial, that is  $U^{n+2\,\mu_1-1} \equiv
U^{n+2\,m-1}$. This in turn is sufficient to estimate the limit
(\ref{limVF}). In fact,
\[
\frac{G\left(U, W(U)\right)}{\Delta(U)}=\frac{G\left(U,
W(U)\right)}{\kappa\,U^n}=\frac{A_1\,U^{n+2\,m-1}+... }{\kappa\,U^n}
\sim \frac{A_1}{\kappa}\,U^{2\,m-1}+...
\]
And now we are ready to formulate the main result of this
subsection.

{\bf Proposition 1.}{\it
 Homoclinic loop bi-asymptotic to stationary
point $(0,\,0)$ of system (\ref{factors2}) corresponds to
compacton-like solution to the initial PDE if $m \le \frac{1}{2} $.
}

Presented above result delivers sufficient but not necessary
condition for the homoclinic loop corresponding to compacton. Basing
on the local asymptotic analysis it is possible to show that for
$0<m<1$ and any natural $n$ trajectory bi-asymptotic to stationary
point $(0,\,0)$ will also correspond to compactly--supported
traveling wave. The proof is somewhat cumbersome  and we shall not
present it here.

We would like to know, what sort of TW corresponds to the homoclinic
loop in case when $m\geq 1$ and $n\geq 1$. Presently we are not able
to answer this question rigorously. Nevertheless, we bring some
arguments evidencing that the ''tail'' of TW in this case spreads up
to $-\infty$ whereas the front sharply ends.  This is so because the
''tail'' corresponds to the outgoing separatrix which is tangent to
either horizontal axis (if $m>1$) or the line $W=\frac{U_1}{V}\,U$
(if $m=1$). On the other hand, the incoming separatrix enters the
origin at the angle $\frac{2\,\pi}{3}$. From these observations
conclusions can be made concerning the shape of TW. Corresponding
arguments will be delivered for the case $m=n=1$.

Analysis of formula (\ref{ALaux2}) enables to state that outgoing
separatrix is tangent to the straight line $W=\frac{U_1}{V}\,U$. So
the first coordinate of the saddle separatrix in vicinity of the
origin satisfies the equation
\[
\frac{d\,U}{d\,\xi}=\frac{U_1}{V}\,U+O\left( U^2  \right),
\]
having the approximated solution $U(\xi)\cong A
exp(\frac{U_1}{V}\,\xi)$. This form of approximate solution tells us
that separatrix reaches the origin as $\xi \rightarrow -\infty$.

Now let us consider the incoming separatrix, which is  tangent to
the vertical axis and lies at the fourth quadrant. This enables us
to assume that $W=-B\,U^{\sigma}+o\left(U^{\sigma}\right)$, where
$0<\sigma<1$ and $B>0$. Approximate equation describing the first
coordinate of incoming separatrix is then as follows
\[
\frac{d\,U}{d\,\xi}=-B \,U^\sigma +o\left(U^{\sigma}\right).
\]
Hence
\[
U(\xi)\cong \left[B (1-\sigma)\,\left(\xi_0-\xi   \right)
\right]^{\frac{1}{1-\sigma}}.
\]
Analysis of this formula shows that trajectory reaches reaches the
origin in finite ''time''.

\section{Results of numerical simulation}\label{sec:numexp}

Numerical simulation of system (\ref{factors2}) were carried out for
different values of the parameters. Let us consider first the
results of simulation obtained for $\kappa=n=1, U_1=3 $ and
$m=\frac{1}{2}$.  Experiments show that at $V$ slightly less then
$V_{cr}=U_1$ a stable limit cycle is softly created in system
(\ref{factors2}).

\begin{figure}
\includegraphics[width=2.5 in, height=2.5 in]{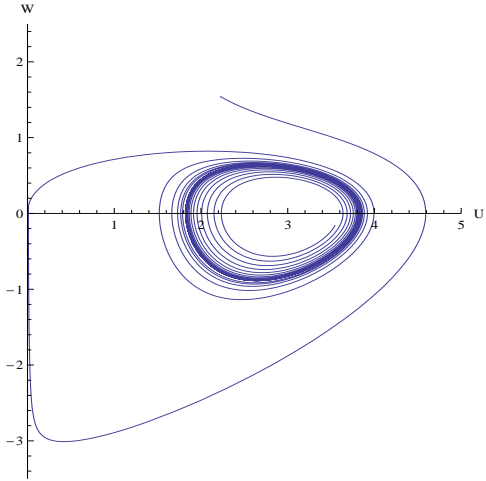}
\caption{Phase portrait of system (\ref{factors2}) obtained for
$\kappa=n=1, U_1=3$, $m=\frac{1}{2}$ and
 $V=2.82$ }\label{fig:Limc1}
\end{figure}

\begin{figure}
\includegraphics[width=2.5 in, height=2.5 in]{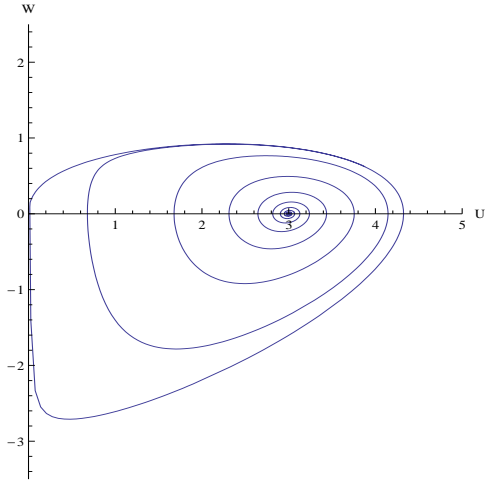}
\caption{Homoclinic bifurcation corresponding to $\kappa=n=1,\,\,
U_1=3$, $m=\frac{1}{2}$ and $V=2.5397975$}\label{fig:Hom1}
\end{figure}

\begin{figure}
\includegraphics[width=2.5 in, height=1.5 in]{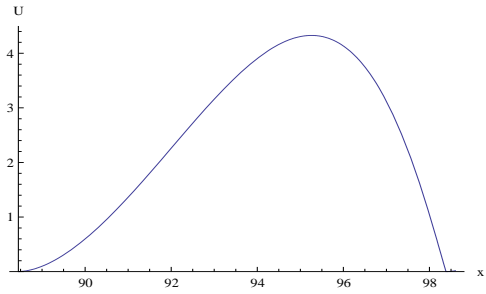}
\caption{Compactly-supported solution of the equation (\ref{GBE1})
corresponding to $\varphi(u)=u^{\frac{1}{2} }$, $\kappa=n=1,\,\,
U_1=3$
 }\label{fig:TW1}
\end{figure}

Its radius grows as $V$ decreases. As the left end of the periodic
trajectory approaches the origin, it  becomes asymmetric (fig.
\ref{fig:Limc1}). At $V \approx 2.5397975$ periodic trajectory
destroys giving way to the homoclinic bifurcation
(fig.~\ref{fig:Hom1}). Corresponding compacton-like solution to the
equation (\ref{GBE1}) is shown on fig.~\ref{fig:TW1}. Compacton,
moving from left to right, is gently sloping towards the tail. It is
worth noting that the compacton-like solution is always symmetric in
case it is solution to Hamiltonian system.

\begin{figure}
\includegraphics[width=2.5 in, height=2.5 in]{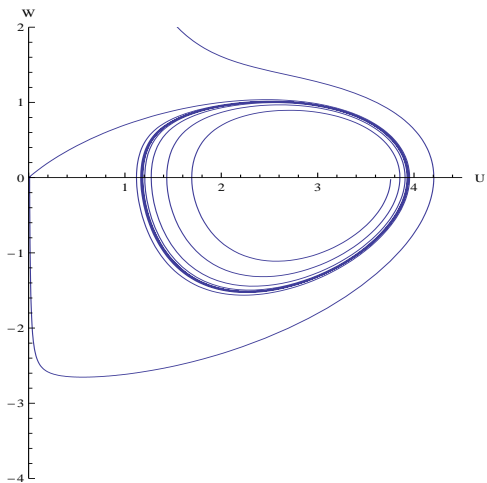}
\caption{Phase portrait of system (\ref{factors2}) obtained for
$\kappa=m=n=1,\,\, U_1=3$ and
 $V=2.775$ }\label{fig:Limc2}
\end{figure}

\begin{figure}
\includegraphics[width=2.5 in, height=2.5 in]{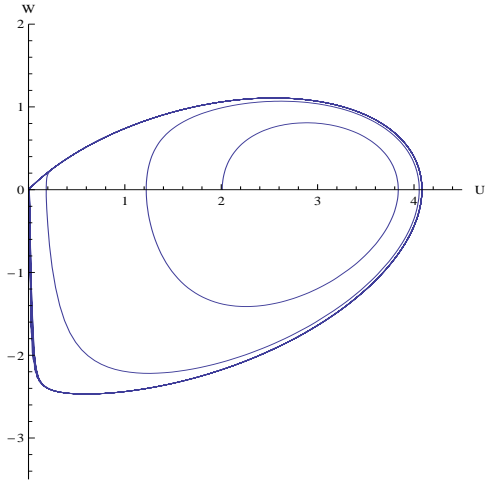}
\caption{Homoclinic bifurcation corresponding to $\kappa=m=n=1,\,\,
U_1=3$ and $V=2.48225$}\label{fig:Hom2}
\end{figure}

\begin{figure}
\includegraphics[width=2.7 in, height=1.5 in]{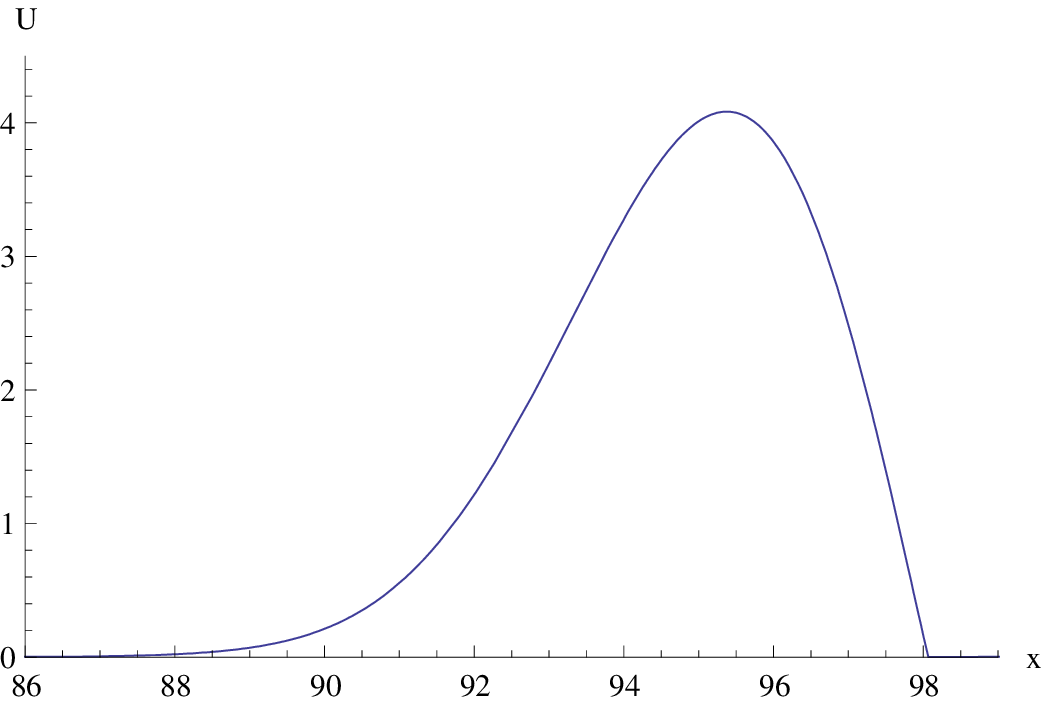}
\caption{Solitary wave solution of the equation (\ref{GBE1})
corresponding to $\varphi(u)=u$, $\kappa=n=1,\,\, U_1=3$ and and
$V=2.48225$ }\label{fig:TW2}
\end{figure}

Another numerical simulations were put up for
$\kappa=1,\,\,m=1,\,\,n=1 ,\,\,U_1=3. $ Scenario is the same as in
the previous case. Limit cycle appearing at   $V<U_1$  grows as $V$
decreases (fig.~\ref{fig:Limc2}). Periodic trajectory is  destroyed
at $V \cong 2.48225$, giving way to homoclinic  trajectory
(fig.~\ref{fig:Hom2}). Corresponding TW solution to source equation
(\ref{GBE1}) is shown on fig.~\ref{fig:TW2}.

\begin{figure}
\includegraphics[width=2.5 in, height=2.5 in]{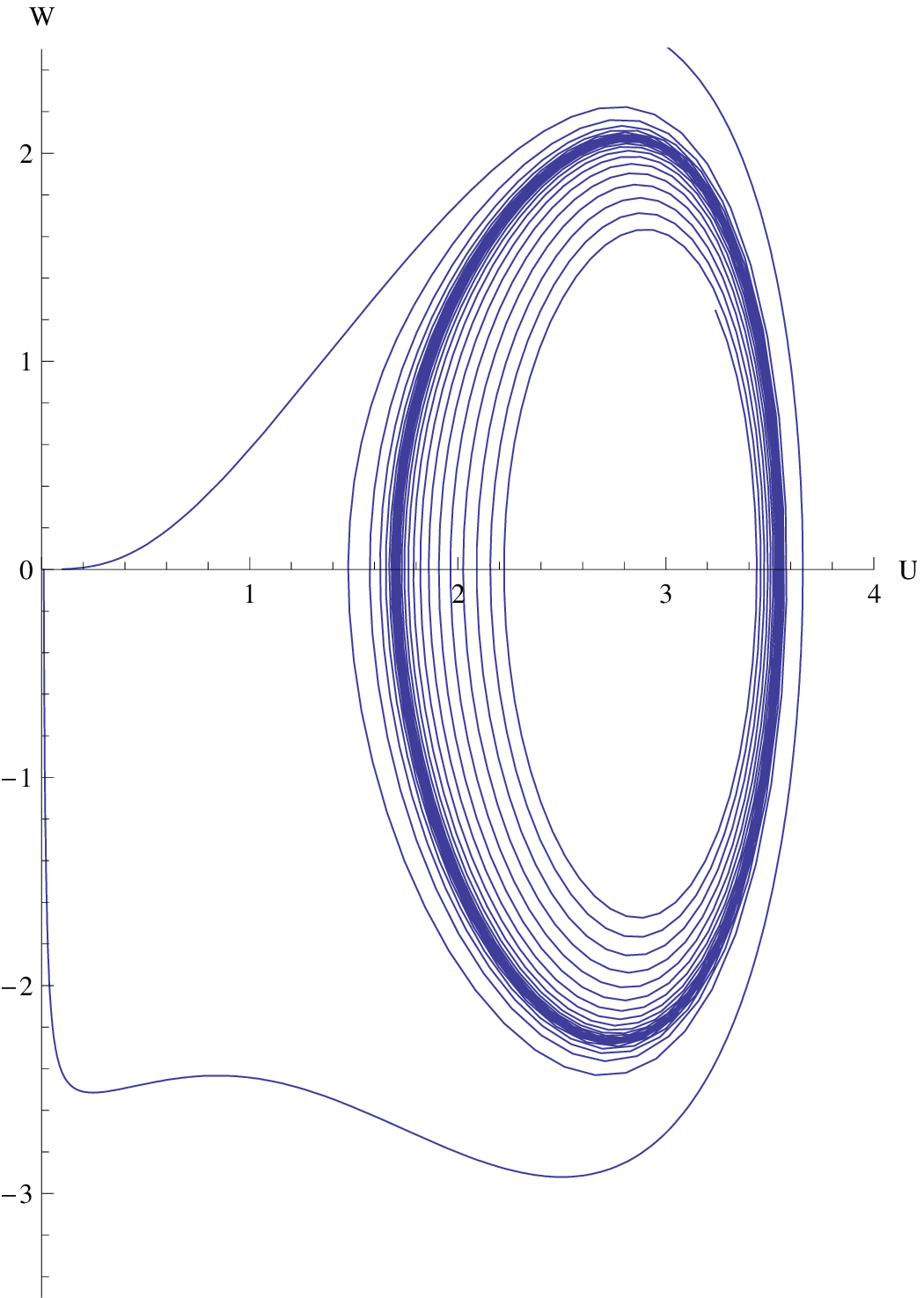}
\caption{Phase portrait of system (\ref{factors2}) obtained for
$m=3,\,\,\kappa=n=1,\,\, U_1=3$ and
 $V=2.7444875$ }\label{fig:Limc3}
\end{figure}

\begin{figure}
\includegraphics[width=2.5 in, height=2.5 in]{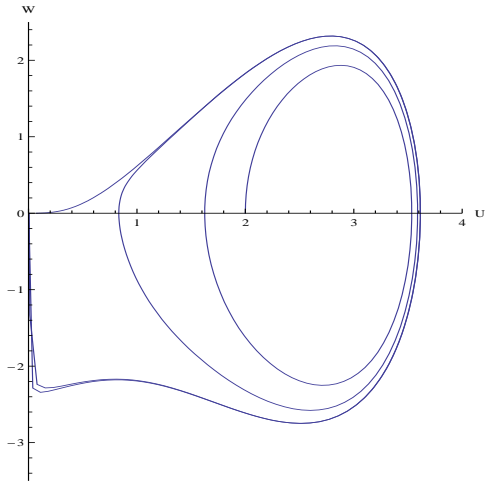}
\caption{Homoclinic bifurcation corresponding to
$m=3,\,\,\kappa=n=1,\,\, U_1=3$ and $V=2.4449$ }\label{fig:Hom3}
\end{figure}

\begin{figure}
\includegraphics[width=2.7 in, height=1.5 in]{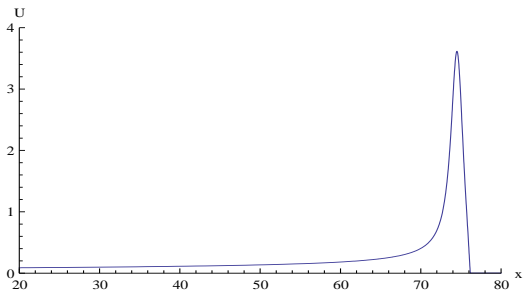}
\caption{Solitary wave solution of the equation (\ref{GBE1})
corresponding to $\varphi(u)=u^3$, $\kappa=n=1,\,\, U_1=3$ and and
$V=2.4449$ }\label{fig:TW3}
\end{figure}

Let us finally show  patterns corresponding to  stronger
nonlinearity of the source term. We take  ${m=3}$  and the remaining
parameter the same as in the previous case. Such a change leads to
larger asymmetry  of the homoclinic contour,
fig.~\ref{fig:Limc3}--\ref{fig:Hom3} and appearance of wave pattern
with sharp front and very long relaxing tail, reminding shock or
detonation wave, fig.~\ref{fig:TW3}.

\section{Final remarks}

So we have shown that, depending on the values of the parameters
convection-reaction-diffusion model described by equation
(\ref{GBE1}) possesses compacton-like TW solutions or mixed
solutions having the soliton-like tails and sharp fronts.  The mere
existence of compacton-like solutions is possible if the diffusion
coefficient is a function of  dependent variable $u$. In all cases
considered wave patterns have been obtained as the result of two
bifurcation. First we performed the local nonlinear analysis of
dynamical system (\ref{factors1}) describing the whole set of TW
solution and stated conditions making possible the stable limit
cycle creation in proximity of critical point $(U_1, 0)$.  The other
critical point, placed at the origin, is shown to possess
simultaneously a saddle sector in the right half-plane, making
possible the homoclinic loop appearance. Next we studied numerically
the evolutions of periodic solutions, observing how they undergo the
homoclinic bifurcation. Analytical results obtained in subsection
~~\ref{asymptorigin} enable us to state when the homoclinic loop
obtained in numerical experiments correspond to compacton.

The results obtained here are not completely rigorous. In particular
it is rather unknown the precise value of wave pack velocity $V$
corresponding to the homoclinic bifurcation.

To obtain the solutions in question analytical methods based on the
generalized symmetries \cite{Olv} and related to them anasatz-based
methods \cite{Fan,OlVor,Vladku_04,Vladku_06} can be applied. Yet it
is little chances to obtain them for all possible values of the
parameters for the source equation is not completely integrable. So
it would be desired to apply some other means, such as the rigorous
computer-assisted proofs for this problem.

In connection with the results obtained a natural question arises
about the benefit of solutions existing merely at selected values of
the parameters. So first let us mention that the velocity of wave
pack, being chosen as the bifurcation parameter, is an ''external''
in some sense parameter and its value is rather connected with the
portion of ''energy'' delivered at the initial moment of time, which
is easily controlled, while the parameters characterizing system
remain unchanged.

Of prime importance is also the well known fact that invariant wave
patterns such as kinks, solitons, compactons etc., {\it very often}
play role of asymptotics, attracting in the long range close near-by
solutions \cite{Barenblatt,Selfsim05,Vlad_08}. Study of the
stability and attracting properties of the invariant solutions to
(\ref{GBE1}) is very non-trivial issue going far beyond the scope of
this article.

\end{document}